\documentclass[a4paper]{article}
\usepackage{amsmath}
\usepackage{amssymb}
\usepackage{color}
\usepackage{graphicx}
\usepackage{rotating}
\usepackage{authblk}

\begin{document}

\title{The shape of the cosmic ray proton spectrum}

\author[1]{\normalsize Paolo Lipari}
\author[2]{\normalsize Silvia Vernetto}

\affil[1]{\footnotesize INFN, Sezione Roma ``Sapienza'',
Piazzale Aldo Moro 2, 00185 Roma, Italy}
\affil[2]{\footnotesize OATO--INAF, INFN Sezione Torino,
Via Pietro Giuria 1, 10124 Torino, Italy}

\date{\small 4th November 2019}

\providecommand{\keywords}[1]{\textbf{\textit{Keywords---}} #1}

\maketitle



\begin{abstract}
Recent observations of cosmic ray protons in the energy range
$10^2$--$10^5$~GeV have revealed that the spectrum cannot be
described by a simple power law.
A hardening of the spectrum around an energy of 
order few hundred~GeV, first observed by the magnetic spectrometers
PAMELA and AMS02, has now been confirmed by several calorimeter detectors
(ATIC, CREAM, CALET, NUCLEON and DAMPE). 
These new measurements reach higher energy
and indicate that the hardening corresponds to
a larger step in spectral index than what estimated by the magnetic spectrometers.
Data at still higher energy (by CREAM, NUCLEON and DAMPE)
show that the proton spectrum undergoes a marked softening at $E \approx 10^4$~GeV.
Understanding the origin of these unexpected spectral features 
is a significant challenge for models of the Galactic cosmic rays.
An important open question is whether additional features are present
in the proton spectrum between the softening and the ``Knee''. 
Extensive Air Shower detectors, using unfolding procedures that
require the modeling of cosmic ray showers in the atmosphere,
estimated the proton flux below and around the Knee (at $E \simeq 3$~PeV).
These results however have large systematic uncertainties
and are in poor agreement with each other.
The measurement in the PeV energy range,
recently presented by IceTop/IceCube, indicates a proton flux
higher than extrapolations of the direct measurements calculated
assuming a constant slope, and therefore requires the existence
of an additional spectral hardening below the Knee.
A clarification of this point is very important for an understanding of the
origin of the Galactic cosmic rays, and is also essential for a precise
calculation of the spectra of atmospheric neutrinos in the energy range
($E \gtrsim 10$~TeV) where they constitute the foreground
for the emerging astrophysical $\nu$ signal.
\end{abstract}

\section{Introduction}
\label{sec:intro} 
The precise measurement of the cosmic ray (CR) spectra
is a fundamental task for high energy astrophysics, because
they encode important information about their sources
and the properties of their propagation in the Galaxy.
In first approximation, the spectra of protons and primary nuclei,
in a broad energy range that extends from approximately 30~GeV
to the so called ``Knee'' at energy few PeV's, have a power--law form.
Recent observations
have however revealed that the spectra deviate from this simple form,
and contain additional structures. In particular, 
direct measurements of the proton flux show the existence of two spectral features:
a hardening at $E \approx 500$~GeV, and a softening at $E \approx 10^4$~GeV.

In this work we review the evidence for these
features and discuss the possible existence of additional
spectral structures between 10~TeV and the Knee,
extrapolating the results of the direct measurements
of the proton flux to higher energy, 
where estimates of the flux have been obtained by 
Extensive Air Shower (EAS) detectors.

Our discussion in the present work will be limited to protons.
This is because they are the largest component
of the CR flux and their spectral shape is measured more accurately.
The comparison of the spectra of different particle types
is of course of great interest. 
Several models for the origin of Galactic cosmic rays predict
that the spectra of protons and primary nuclei
(such as helium, carbon, oxygen, ...)
have, at least in good approximation,
the same spectral shape when expressed as a function of rigidity,
and it is clearly very important to study experimentally
the validity of this assumption.
A discussion of the spectra of CR nuclei will be however
postponed to a future publication.

This work is organized as follows. In Sec.~\ref{sec:direct} 
we analyze the direct (magnetic and calorimetric)
measurements of the $p$ flux above 100~GeV, and discuss
the evidence for the existence and the properties of the two
(hardening and softening) features. 
In the following section we compare the extrapolations
of a fit to the direct measurements with the published
estimates of the proton spectrum obtained by EAS detectors.
In a final section we summarize the results and give some conclusions.

\section{Measurements of the CR proton spectrum}
\label{sec:direct}
The existence of a hardening in the energy spectrum of protons
(and other CR species)
was first inferred in a paper of the CREAM Collaboration \cite{Ahn:2010gv}
from a comparison of their measurements of the spectral indices at 
$E \gtrsim 1$~TeV, with the results obtained by magnetic spectrometers
at lower energy ($E \sim 100$~GeV). The flatter spectra observed at
high energy suggested the existence of a ``discrepant hardening'' in
the energy range (0.2--1~TeV) where data were not yet
available.

Hardenings in the proton and helium spectra were 
then directly observed by PAMELA \cite{Adriani:2011cu},
that reported the existence of ``sharp'' breaks at rigidity 
($\rho_p = 232^{+35}_{-30}$~GV for protons and $\rho_{\rm He} \simeq 243^{+27}_{-31}$~GV for helium).
At the break rigidity the proton flux
was observed to flatten from a spectral index
$\alpha_1 \simeq 2.85\pm 0.04$
to an index $\alpha_2 \simeq 2.67\pm 0.06$.
These results generated considerable interest because
simple and commonly accepted models
for Galactic cosmic rays predicted spectra
with the shape of an unbroken power law
(above the maximum energy where solar modulations have a significant effect and
up to the Knee at $E \simeq 3$~PeV).

The existence of the hardenings for protons and helium nuclei
was then confirmed, with higher statistical significance, 
by the AMS02 detector \cite{Aguilar:2015ooa,Aguilar:2015ctt}.
The spectral hardening for protons measured by AMS02
is centered at higher rigidity
($\rho_p \simeq 336^{+95}_{-52}$~GV)
in comparison to the PAMELA results,
and has a smaller step in spectral index
$\Delta \alpha \simeq 0.133^{+0.056}_{-0.037}$
(with $\alpha_1 \simeq 2.849^{+0.006}_{-0.005}$, and $\alpha_2 \simeq 2.716^{+0.037}_{-0.056}$).

Measurements of the proton spectrum above 1~TeV have been
now obtained by several CR calorimeters:
ATIC \cite{Panov:2011ak}, 
CREAM \cite{Yoon:2017qjx}, 
CALET \cite{Adriani:2019aft},
NUCLEON \cite{Atkin:2018wsp,Atkin:2019xxt}
and DAMPE \cite{An:2019wcw}.
The ATIC, CREAM and NUCLEON detectors cover only 
the energy range $E \gtrsim 1$~TeV,
and therefore do not observe directly the hardening,
but only the spectral shape above it. 
The CALET and DAMPE instruments measure the proton spectrum across
the entire region of the hardening.

Three detectors (CREAM, NUCLEON and DAMPE)
reach a maximum energy of order 100~TeV,
and they all have indications
(or clear evidence in the case of DAMPE) for the existence
of a second spectral feature: 
a softening centered at an energy of order 10~TeV.

\begin{figure}[bt]
\begin{center}
\includegraphics[width=17.0cm]{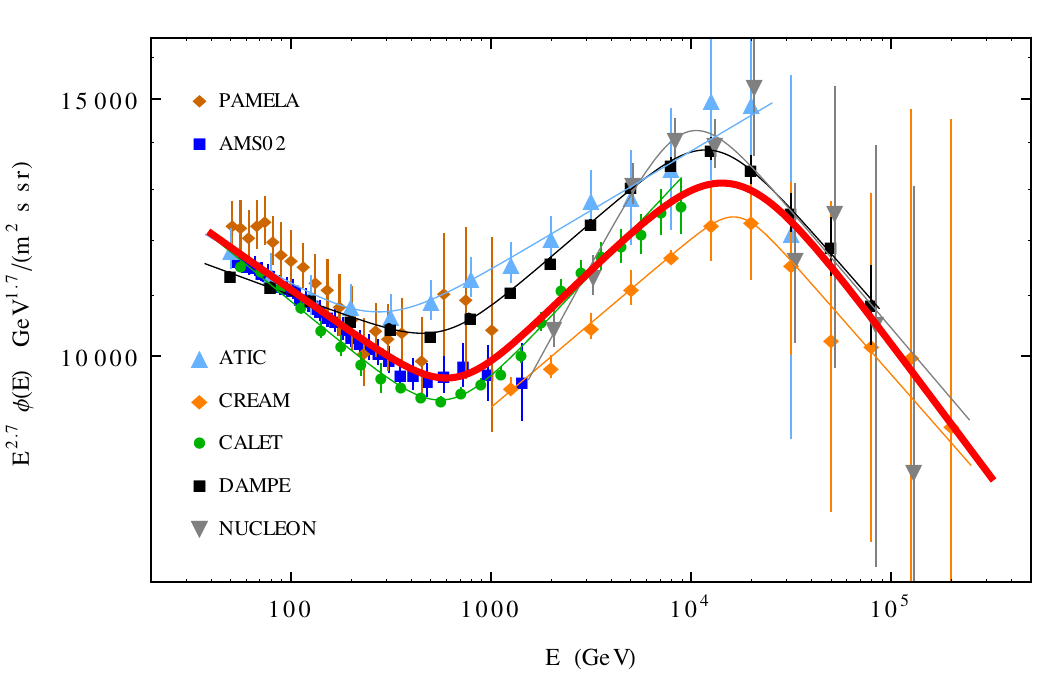}
\end{center}
\caption {\footnotesize
Direct measurements of the CR proton spectrum.
The flux is shown in the form
$E^{2.7} \, \phi (E)$ versus $E$ to enhance the visibility
of the spectral features.
The points are the data of
PAMELA \protect\cite{Adriani:2011cu},
AMS02 \protect\cite{Aguilar:2015ctt},
ATIC \protect\cite{Panov:2011ak}, 
CREAM \protect\cite{Yoon:2017qjx},
CALET \protect\cite{Adriani:2019aft}, 
DAMPE \protect\cite{An:2019wcw}
and NUCLEON \protect\cite{Atkin:2018wsp}.
The thick (red) solid line is a fit of the
combined data of all the experiments
using the two--break expression (\ref{eq:fit2}). 
The thin lines are fits of the data of individual experiments.
The parameters of all fits are listed in Table~\ref{tab:par}.
\label{fig:prot_direct} }
\end{figure}

The measurements of the proton flux listed above are shown in
Fig.~\ref{fig:prot_direct} plotted in the form
$E^{2.7} \, \phi(E)$ versus the kinetic energy $E$ to enhance
the visibility of the spectral features.
The error bars of the PAMELA and AMS02 data points shown in the figure 
are calculated combining quadratically statistical and systematic errors.
For the other detectors they only show statistical uncertainties.
For CREAM, CALET and DAMPE the systematic error
(mainly due to the calibration of the energy scale) is of order 10\%.
For ATIC we estimated a systematic error of 20\%.
For the NUCLEON detector the spectrum given in the figure
is the one obtained estimating the energy with the 
kinematical method KLEM (Kinematic Lightweight Energy Meter),
and the systematic error is of order 25\%.

Taking into account the systematic errors,
all the measurements in Fig.~\ref{fig:prot_direct}
are in reasonable good agreement in the energy ranges
where they overlap (see more discussion below).
A visual inspection of the figure immediately shows that
a simple power--law cannot give a good description of
the observations, and both the hardening (at $E \approx 500$~GeV)
and the softening (around $E \approx 10$--20~TeV) are clearly visible.

To study more quantitatively the spectral shape of the proton flux
we have fitted the data using the ``two--break'' expression:
\begin{equation}
 \phi(E) =
 K~\left ( \frac{E}{E_0} \right)^{-\alpha_1} ~
 \left [1 + \left (\frac{E}{E_b} \right )^{1/w} \right ]^{-(\alpha_2 -\alpha_1)\, w} ~
 \left [1 + \left (\frac{E}{E_b^\prime} \right )^{1/w^\prime} \right ]^{-(\alpha_3 -\alpha_2)\, w^\prime} 
\label{eq:fit2}~
\end{equation}
where $E_0$ is an arbitrary reference energy,
$K$ gives the absolute normalization, and
$\{\alpha_1, E_b, \alpha_2, w, E_b^\prime, \alpha_3, w^\prime \}$
are free parameters that determine the spectral shape.
The quantities $E_b$ and $E_b^\prime$ are the ``break'' energies
around which the spectral index changes,
all other parameters are adimensional.
Without loss of generality one can impose the constraints
$E_b < E_b^\prime$, and $w$ and $w^\prime$ positive.

Expression (\ref{eq:fit2}) is an obvious generalization of the ``one--break'' form:
\begin{equation}
 \phi(E) =
 K~\left (\frac{E}{E_0} \right )^{-\alpha_1} ~
 \left [1 + \left (\frac{E}{E_b} \right )^{1/w} \right ]^{-(\alpha_2 -\alpha_1)\, w} 
\label{eq:fit1a}
\end{equation}
that describes a flux with a single spectral feature centered
at the energy $E_b$, where the spectral index changes from 
$\alpha_1$ (for $E \ll E_b$) to $\alpha_2$ (for $E \gg E_b$). The parameter
$w$ gives the width in $\log E$ of the step in spectral index
with $w =0$ corresponding to a discontinuous variation.
The relation between the parameter $w$ and the width in $\log E$
is discussed in detail in \cite{Lipari:2017jou}.
A useful expression for this relation is:
\begin{equation}
 \left (\Delta \log_{10} E \right )_{\Delta \alpha/2}
 = (\log_{10}9) \; w 
 = 0.954 \; w ~.
\label{eq:w}
\end{equation}
This equation states that one half of the step in spectral index
 develops in an interval of $\log_{10} E$ approximately equal to $w$.

In the two--break expression of Eq.~(\ref{eq:fit2}) the
spectrum has slope $\alpha_1$ for $E \ll E_b$ and
$\alpha_3$ for $E \gg E_b^\prime$, while in the interval 
$E_b < E < E_b^\prime$ the spectral index is approximately
equal\footnote{
This is the case when the ratio $E_b^\prime/E_b$ 
is sufficiently large in comparison to the
widths of the two features. That is when
$\log_{10} (E_b^\prime /E_b) \gtrsim (w + w^\prime)$.}
to $\alpha_2$.

The generalization of expression (\ref{eq:fit2})
for a spectrum that contains an arbitrary number of
features is obvious, and simply requires the inclusion
of a new distortion factor for each feature.
Each factor (in square parenthesis in the formula) depends on 
three parameters $\{E_{b,j}, \Delta \alpha_j, w_j\}$ that give the position,
size and width of the $j$--th feature.

The family of curves described by Eq.~(\ref{eq:fit1a})
has been used to fit ``features'' in the CR spectra
(including the proton flux) in several papers,
using however different parametrizations.
The main source of potential confusion is the 
choice for the parameter that describes
how ``gradual'' is the step in spectral index.
AMS02 \cite{Aguilar:2015ooa},
CALET \cite{Adriani:2019aft} and DAMPE \cite{An:2019wcw}
adopted the parameter $s = w\, (\alpha_2-\alpha_1)$, while 
NUCLEON \cite{Atkin:2018wsp} uses
the parameter $S = 1/w$.
The choice made here is motivated by the fact
that the parameter $w$ has a simple and clear physical meaning
as the width of the interval in $\log E$ where the spectral index undergoes its step,
independently from the value of $\Delta \alpha$ [see Eq.~(\ref{eq:w})].

To study the shape of the proton spectrum and investigate how
consistent are the different measurements we have performed
separate fits to the data of the individual experiments.
The DAMPE data \cite{An:2019wcw}
cover an energy interval that includes both spectral
features, and have been fitted with the two--break expression of Eq.~(\ref{eq:fit2}).
The data of all other detectors
cover an energy interval that includes only one of the two features
(the lower energy hardening for PAMELA, AMS02, ATIC and CALET,
and the higher energy softening for CREAM and NUCLEON),
and have then been fitted with the single break
expression of Eq.~(\ref{eq:fit1a}), using the appropriate and obvious
renaming of the parameters when needed (for CREAM and NUCLEON).

The best fit spectra are shown in
Fig.~\ref{fig:prot_direct} and the parameters of the fits are listed in Table~\ref{tab:par}.
Our results are consistent with the fits reported in the original publications.

The fits to the PAMELA and AMS02 data are performed
combining quadratically statistical and systematic uncertainties.
For the other experiments we use only 
statistical uncertainties.
In these experiments the systematic errors for different points are strongly correlated
and the fits performed with only statistical uncertainties
give very good values for $\chi^2_{\rm min}/N_{\rm dof}$ (see Table~\ref{tab:par})
(while fits using also systematic errors and neglecting correlations
gives too small values of order 0.01--0.03).

Three data sets (DAMPE, NUCLEON and CREAM) show evidence for
the existence of a spectral softening and weaker indications
for this feature are also present in the ATIC and CALET data.

It should be noted that the CREAM Collaborations in
\cite{Yoon:2017qjx} does not claim the detection of
a softening feature, and fits the data
taken in the energy interval [1,200]~TeV
with a simple unbroken power law obtaining a 
the spectral index $\alpha \simeq 2.61 \pm 0.01$.
The CREAM paper discusses the possibility of a deviation
from the power law form stating:
``The spectra become softer above $\sim 20$~TeV.
However, our statistical uncertainties are large at these energies
and more data are needed''.
It is nonetheless interesting to investigate quantitatively how strong is the
indication of a softening.
Fits of the CREAM data (that use only statistical errors)
for a simple power law and with the one--break expression of Eq.~(\ref{eq:fit1a})
gives $\chi^2_{\rm min} \simeq 13.3$ and 1.9 respectively.
Taking into account the different number of free parameters in the two cases,
this corresponds to an indication with a significance of order 3.0~$\sigma$ in
favor of the existence of a softening.

We also performed a global fit of the data of all seven experiments
in the energy range $E > 50$~GeV using expression (\ref{eq:fit2}).
The resulting best fit
(obtained combining quadratically statistical and systematic uncertainties) 
is shown (as a thick red line) in Fig.~\ref{fig:prot_direct},
and the parameters are given in the right--hand column of Table~\ref{tab:par}.
The fit is of good quality, with $\chi^2_{\rm min} = 46.5$ for 116 d.o.f.

From this study we can draw the following conclusions:
\begin{itemize}
\item In the energy range below the hardening the $p$ spectrum
 has been very accurately measured by the magnetic spectrometers
 AMS02 and PAMELA that obtain best fits to
 the spectral index that are nearly identical
 ($2.850 \pm 0.043$ for PAMELA
 \cite{Adriani:2011cu}
 and $2.849^{+0.006}_{-0.005}$ for AMS02 \cite{Aguilar:2015ooa}).
 The data of the calorimeter experiments (ATIC, CALET and DAMPE)
 are consistent with these results, even if the best fit of DAMPE
 indicates a slightly harder spectrum
 (with a slope given as $2.772\pm 0.002$
 in \cite{An:2019wcw}, while our fit yields $2.750 \pm 0.005$).

\item The hardening of the spectrum for an energy (or rigidity)
 of order few hundred GeV is now very well established, but the
 precise shape of the feature is not well determined because
 the measurements of different experiments are not in perfect
 agreement with each other.
 The uncertainties are associated to all the three parameters
 that describe the shape of the feature: break energy,
 spectral index step $\Delta \alpha$ (or equivalently the
 slope $\alpha_2$ above the hardening) and width $w$.
 The slope $\alpha_2$ of the proton spectrum
 in the energy range 1--10~TeV has now
 been measured by five different calorimeter experiments
 (ATIC, CREAM, CALET, DAMPE, and NUCLEON).
 For all these experiments the spectrum in the 1--10~TeV energy range 
 is harder ($\alpha_2$ between 2.44 to 2.62)
 than what is indicated by the AMS02 data
 ($\alpha_2 \simeq 2.716^{+0.037}_{-0.056}$ \cite{Aguilar:2015ooa}).
 The most natural explanation for this discrepancy is that the
 AMS02 data, that are limited to a maximum rigidity of 1.2~TV,
 do not measure well the entire ``development'' of the hardening.
 Our fit to the combined data of all experiments
 yields a good quality fit ($\chi^2_{\rm min}/N_{\rm dof} \simeq 0.40$)
 with a gradual ($w \simeq 0.27 \pm 0.19$)
 hardening centered at $E_b \simeq 650^{+260}_{-150}$~GeV
 with a spectral index at high energy $\alpha_2 = 2.57^{+0.04}_{-0.06}$.

\item The spectrum of protons exhibits a second
 spectral feature, a softening at $E \simeq 10$--20~TeV.
 This conclusion is now supported by the observations of three experiments:
 CREAM, NUCLEON and DAMPE
 (with weaker indications in the data of ATIC and CALET).
 The evidence of DAMPE is the strongest, with a statistical significance
 estimated as 4.7~$\sigma$ in \cite{An:2019wcw}.
 the simpler analysis performed by us using only the statistical errors
 gives a very large $\Delta \chi^2 = 143$ between the two--break
 fit shown in Fig.~\ref{fig:prot_direct} and one--break fit
 that extends as a simple power law above the hardening
 (corresponding nominally to more than 11~$\sigma$).
 For the CREAM and NUCLEON data the existence of a spectral
 softening has a statistical significance of order 3~$\sigma$.
\end{itemize}

\section{Air Shower measurements of the proton flux}
\label{sec:EAS}
A question that emerges immediately from our discussion on the
direct measurements of the CR proton flux is the shape of the
spectrum at higher energy (in the range 10--$10^3$~TeV).
Does the spectrum continues as an unbroken power law
up to the Knee or does it contain additional structures?

At present, direct measurements of the CR spectra 
extend only to a maximum energy of order $E \simeq 100$~TeV, but
information on the spectra at higher energy 
can be obtained from observations of the Extensive Air Showers (EAS)
generated by primary cosmic rays in the Earth's atmosphere.
These measurements are however
model dependent because they require 
a comparison of the data with Montecarlo simulations of the shower development.
The main uncertainty arises from our poor knowledge of the properties
of hadronic interactions
(such as multiplicity distributions and inclusive energy spectra
of the particles in the final state of a collision).

\begin{figure}[bt]
\begin{center}
\includegraphics[width=17.0cm]{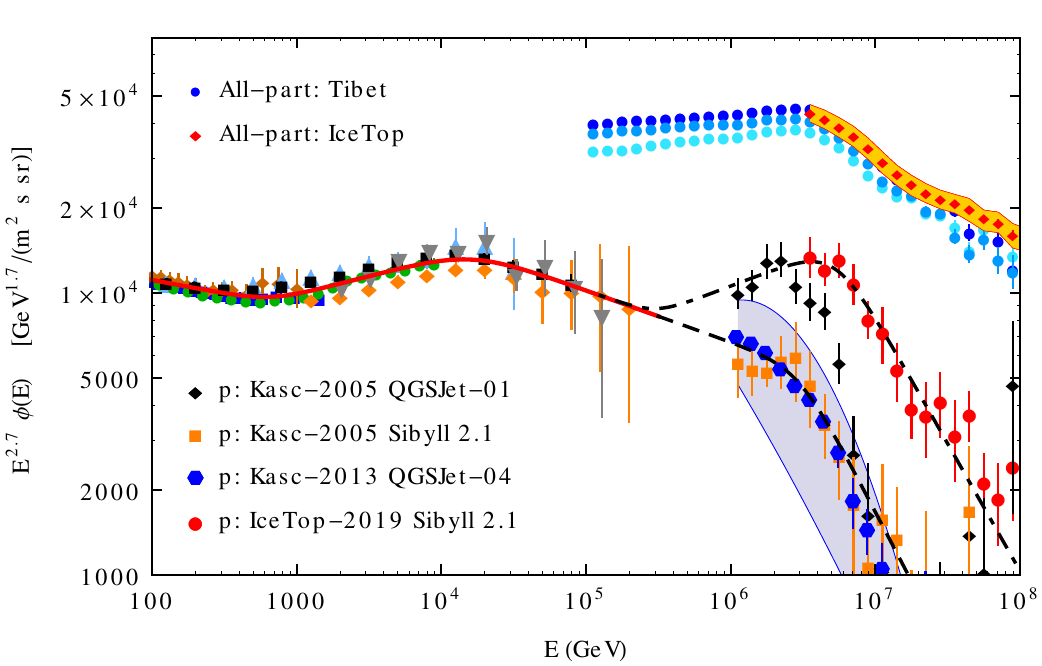}
\end{center}
\caption {\footnotesize
All--particle and proton spectra 
obtained by direct measurements and EAS observations.
The all--particle data are by the Tibet experiment
\protect\cite{Amenomori:2008aa} (with three sets of data points
obtained with different assumptions for the CR composition and shower
development models), and by IceTop/IceCube \protect\cite{IceCube:2019hmk}
(with the shaded area indicating systematic uncertainties).
For the proton direct measurements the symbols are identical
to those in Fig.~\ref{fig:prot_direct}.
The EAS proton spectra are by Kascade--2005 \protect\cite{Antoni:2005wq},
Kascade--2013 \protect\cite{Apel:2013uni}
(with the shaded area indicating systematic uncertainties)
and IceTop/IceCube--2019 \protect\cite{IceCube:2019hmk}.
The thick solid line is a fit to the direct measurements of the proton flux
(with the parameters given in Table~\protect\ref{tab:par}).
The dashed and dot--dashed lines are extrapolations
to higher energy (see main text).
\label{fig:prot_eas}}
\end{figure}

Several experiments have published results on the all--particle spectrum,
that sums the contributions of all CR components
at the same energy per particle:
\begin{equation}
\phi_{\rm all} (E) = \sum_{\rm A} \phi_{\rm A} (E) ~.
\end{equation}
Fig.~\ref{fig:prot_eas} shows the all--particle spectrum
measured by the Tibet detector \cite{Amenomori:2008aa}
in the energy range $10^5$--$10^7$~GeV and
by the IceTop/IceCube detector \cite{IceCube:2019hmk}
in the energy range $10^7$--$10^8$~GeV.
For the Tibet experiment the figure gives three estimates of
the spectrum that differ by approximately $\pm 10\%$,
obtained in \cite{Amenomori:2008aa}
using different assumptions about the CR composition
and different Montecarlo codes to model the hadronic interactions.
For the IceTop/IceCube measurement the estimated systematic error
(shown as a shaded area) is also of order 10\%.
Taking into account these systematic uncertainties, the results
are in good agreeement.
Other measurements of the all--particle spectrum
(for a review see \cite{Gaisser:2013bla}),
including those of Kascade \cite{Antoni:2005wq},
are also consistent. 
The all--particle spectrum around 
$E \approx 3$~PeV exhibits the well--known and much discussed softening
called the Knee, whose origin remains controversial and poorly understood.

The identification of the mass of the primary particlea and the measurement
of the spectra of individual CR components is a more difficult task,
with larger systematic uncertainties.
The published estimates of the proton spectrum obtained by EAS detectors
are shown in Fig.~\ref{fig:prot_eas}.

In 2005 the Kascade Collaboration \cite{Antoni:2005wq}
published an ``unfolding'' of the CR flux
as the sum of five mass components ($p$, He, C, Mg, Fe).
In the Kascade paper the unfolding is performed
using two different Montecarlo codes to model high energy hadronic interactions:
QGSJet--01 \cite{Kalmykov:1993qe} and Sibyll 2.1 \cite{Engel:1999db}
(in both cases low energy interactions were modeled using the GHEISHA code
\cite{Fesefeldt:1985yw}).
The proton spectra estimated with these two methods (shown as diamonds and
squares in Fig.~\ref{fig:prot_eas}) have similar shapes but their normalizations
differ by approximately a factor of two for energies below the Knee
($E \simeq 1$--2~PeV), so that the proton component accounts
of approximately 30\% of the all particle flux in one case, and only 15\% in the other.

In 2013 the Kascade--Grande Collaboration
\cite{Apel:2013uni}
published a reanalysis
of the Kascade data based on improved methods, and using as Montecarlo code
for high energy interactions a new version
of the QGSJet code \cite{Ostapchenko:2004ss},
and at low energy the FLUKA code \cite{Battistoni:2007zzb}.
The unfolded proton spectrum is shown as hexagons in Fig.~\ref{fig:prot_eas}.
In this case the proton knee has a more gradual shape with a normalization that
is similar to the one obtained using the Sibyll code in the previous analysis.
The total systematic uncertainty estimated for this analysis \cite{Apel:2013uni}
is shown as the shaded area.

An estimate of the proton flux at very high energy has also been obtained by
the IceTop/IceCube Collaboration \cite{IceCube:2019hmk},
unfolding the spectrum into
four components ($p$, helium, oxygen and iron).
The proton flux from this analysis
is shown in Fig.~\ref{fig:prot_eas} as circles.
In this study, the Montecarlo code used to model the development of
high energy showers is Sibyll 2.1, also used by
Kascade in \cite{Antoni:2005wq}, 
however the normalization of the proton spectrum is now
much higher (for energy around 2~PeV the difference in normalization
is approximately a factor of three).

The reasons for such large differences between estimates
of the proton flux in the PeV energy range are not clear,
and this question certainly deserves a detailed study.

The large uncertainties about the PeV proton flux  make
the extrapolation of the direct measurements ambiguous.
Different extrapolations can be constructed to
agree with PeV data.
This is illustrated in Fig.~\ref{fig:prot_eas} that
shows two possible extrapolations models.

One model (shown as the dashed line) assumes that the
proton spectrum continues with a constant spectral index
above the softening at $E \approx 15$~TeV.
With this assumption the extrapolation is in reasonable
agreement with the estimate of Kascade--2013 \cite{Apel:2013uni}.

A second model (shown as the dot--dashed line) 
is constructed to agree with the estimate of the proton
measurement obtained by IceTop/IceCube.
In order to do so, it is necessary to introduce additional
structures in the spectrum.
To connect the measured proton flux at $E_1 \simeq 20$~TeV
(where good precision direct measurements are available) to
the IceTop/IceCube data at $E_2 \simeq 2$~PeV, one needs an average spectral index
\begin{equation}
\langle \alpha (E_1, E_2) \rangle = -\frac{\ln (\phi_2/\phi_1)}{\ln E_2/E_1}
\end{equation}
Substituting in this expression the measured values
one obtains the estimate
$\langle \alpha \rangle \simeq 2.66 \pm 0.04$,
where the error is calculated assuming a 15\% error on each flux measurement.
This value is significantly smaller
than the slope measured with direct observations
of the proton flux above 15~TeV ($\alpha_3 \simeq 2.87$).
One must then conclude that if the IceTop/IceCube 
(and also the direct) measurements are correct,
then the proton spectrum cannot have a constant
spectral index in the energy interval $[E_1,E_2]$,
and must contain a hardening.
These considerations obviously do not allow to
predict the detailed form of the spectrum, and the
dot--dashed line 
shown in Fig.~\ref{fig:prot_eas} is only one possible example.

\section{Discussion} 
\label{sec:discussion}
Precise measurements of the CR proton flux in the energy
range 10$^2$--$10^5$~GeV have shown that the spectrum
deviates from the simple power law shape predicted by
commonly accepted models and exhibits two features.
The first one is the sub--TeV hardening discovered by PAMELA,
then observed by AMS02 and several calorimeter detectors.
Measurements in the multi--TeV energy range indicate
that the step in spectral index of the hardening is quite large,
with slopes of order 2.85 and 2.60 below and above the break energy.

A second feature, a softening, has now been clearly detected
at an energy of order 10~TeV by DAMPE, confirmining
the indications already present in the CREAM and NUCLEON observations. 
Taking into account the systematic uncertainties the
measurements of these three detectors are in good agreement,
with absolute fluxes that differ by approximately 10\%
around $E \simeq 10$~TeV.
Our global fit gives a break energy with a rather large error
$E_b^\prime = 16^{+13}_{-8}$~TeV.
The slopes measured above the softening are
also consistent (see Table~\ref{tab:par}).
Our fit to all data above the softening
gives a spectral index $2.87^{+0.15}_{-0.10}$.

There is clear evidence that the hardening is also present
in the spectra of helium and other primary nuclei
 \cite{Adriani:2011cu,Aguilar:2015ctt,Aguilar:2017hno},
and the results are consistent with the hypothesis that the feature
has a ``universal'' shape when studied as a function of rigidity
(even if the values of the slopes of
different particle types are not equal). 

The existing measurements of helium and other nuclei at
$E \gtrsim 10$~TeV
\cite{Yoon:2017qjx,Atkin:2018wsp,Atkin:2019xxt}
suggest that the softening is also present for
cosmic ray particles heavier than protons, and within the large 
statistical and systematic errors, the feature is again consistent
with having a rigidity dependent shape, common to all particle types.
More data are however required to reach a firm conclusion.

It is also interesting to note that the
HAWC measurement of the all--particle spectrum
\cite{Alfaro:2017cwx} in the 10--500~TeV energy range
shows a spectral break at the energy $45.7 \pm 0.1$~TeV where
the index is oberved to change from 
$2.49\pm 0.01$ to $2.71\pm 0.01$ (with a step of approximately
the same size as the one observed for the proton flux).
Evidence for a strong break in the all--particle spectrum
at a rigidity 7--17~TV has also been reported by the NUCLEON experiment
\cite{Atkin:2018wsp,Atkin:2019xxt}.

Observations of the ``light component'' (proton and helium)
of the CR spectrum in the energy range 5--200~TeV
have been performed by the EAS detector ARGO--YBJ
\cite{Bartoli:2012zz} reporting a spectrum
consistent with an unbroken power law.
A subsequent measurement of the light component of the CR spectrum
by ARGO--YBJ at higher energy has given indications for an intriguing
``cutoff'' at $E \approx 640\pm 87$~GeV \cite{Bartoli:2015vca}.

For an understanding of the origin of the shape of the proton spectrum
it is very important to establish if additional spectral structures exist
below the Knee.
For this purpose it would be clearly desirable to extend
the direct measurements to higher energy.
At present measurements of the CR spectra 
at very high energy have been obtained by EAS detectors.
This is a very difficult task,
and the results are not in good agreement with each other.
Estimates of the proton flux in the PeV energy range
performed by Kascade and IceTop/IceCube for $E \simeq 2$~PeV
differ by a factor larger than three.

These discrepancies do not allow to extrapolate
the direct measurements of the proton flux in a
unique way. Low estimates of the flux in the PeV energy range
(such as those presented by Kascade in \cite{Apel:2013uni})
are consistent with the hypothesis that the $p$ spectrum continues
with a constant index from the softening break at $\sim 15$~TeV up to the
Knee. On the other hand, the higher flux recently obtained
by IceTop/IceCube \cite{IceCube:2019hmk} 
requires the existence of additional spectral structures
(with a new hardening).

Future EAS observations by detectors
with the capability of measuring several components
(electromagnetic, muonic, Cherenkov light, $\ldots$)
of the showers (such as LHAASO \cite{Bai:2019khm}) have the potential to
significantly reduce the uncertainties in the
measurement of the spectra for different primary particles.

The precise description of the CR spectra below the Knee
is crucial for the calculation of the spectra of atmospheric neutrinos
in the critical energy range ($E_\nu \sim 10$--100~TeV)
where one observes the transition between a 
flux dominated by the softer atmospheric component
to a flux dominated by astrophysical particles.

The discovery of the softening in the proton spectrum
can have very important consequences for this problem.
The models of the atmospheric neutrino spectra 
that are currently used to study the data of
high energy neutrino telescopes are
calculated using the assumption \cite{Engel:1999db}
that the primary CR spectra are simple power laws
between the sub--TeV hardening and the Knee.
Introducing the softening feature for primary protons
and extrapolating with a constant slope to high energy (up to the Knee)
results in neutrino spectra that are lower and softer.
A too large reduction of the calculated $\nu$ flux
could in fact be in tension with the IceCube observations.

A quantitative study of this problem
requires the modeling of the spectra not only for protons but also
for all other nuclei that give smaller but important
contributions to the atmospheric neutrino fluxes, and will not
be developed here.
Present and future observations of the neutrino fluxes can however
be used to constraint the primary CR fluxes.

Developing an understanding of the
astrophysical mechanisms that shape
the Galactic CR spectra and generate
the observed features is a challenging undertaking.
The standard ``text--book'' models for the origin
of the Galactic cosmic rays
(see for example \cite{Gaisser:2016uoy})
predict (for protons and primary nuclei) spectra
that have a simple power law shape,
(up to a maximum energy determined by the properties
of the accelerators) with a spectral index that, in good approximation,
is equal for all particle types.
In these models the spectra are constructed
assuming a ``universal'' CR source spectrum
that has power law form with a slope $\alpha_0$
that is distorted by propagation effects
with the rigidity dependence
$\propto \rho^{\delta}$.
The observable spectrum is then also of power law form
with index $\alpha \simeq \alpha_0 + \delta$.
Such a scenario fails to describe the observed spectra.

After the discovery of the sub--TeV hardening
by PAMELA, several authors have constructed
models for the Galactic cosmic rays where the spectra
do have such a spectral feature.
Essentially all of these models are simple
generalizations of the ``standard'' scenario described above. 
In the models the hardening is attributed to
either the acceleration or the propagation of cosmic rays.
In the first case one for example assume
the existence of two different classes of accelerators
that generate source spectra
of different slopes and dominate the spectra below and
above the hardening feature, or alternatively the existence
of two regimes for acceleration at low and high energy.
In the second case one can assume
that the propagation effects have different rigidity
dependences at low and high rigidity.

In all cases, the modified models predict
that above the hardening the CR spectra continue with a
constant slope up to the Knee.
The discovery of the softening is therefore a surprise
that requires the construction of new (and more complex) models.

In principle, to explain the presence of two (and not just one)
spectral features it is still possible to adopt what is essentially
the same ``methodology'' discussed above to explain a
single hardening.
For example one can assume the existence of three
regimes for CR propagation in three different rigidity intervals,
or introduce more classes of accelerators, or a combination of
these effects. 
This approach can also be generalized to the presence of
three of more spectral features, however this
appears more and more unnatural and contrived.

Perhaps, especially if the data will reveal
the existence of more spectral features,
it could become interesting, or in fact necessary,
to look for a different type of explanation.

It is interesting to note that the features in the CR spectra
that we have discussed in this work can be considered as rather
subtle distortions of a flux of power law shape, 
that are revealed only by high precision measurements.
An attractive (even if quite ``non--orthodox'')
possibility is that the CR spectra are generated by
an ensemble of sources that
release in interstellar space spectra that
(in contrast with the ``standard'' scenario)
do not have one (or only few) shapes,
but have a large variety of shapes that individually are not of
power law form. These components could then combine to form an average spectrum
that has a nearly (but not exactly) power law form.
The variations with energy of the spectral index of the observed CR fluxes 
could then reflect the existence of these multiple source components.

In this scenario the production of Galactic cosmic rays
can be considered analogous to the formation of the
time averaged spectrum of solar energetic particles,
generated by the ensemble of solar flares \cite{mewaldt_2007}.
The spectra of the individual flares have a large variety of spectral shapes,
(that in general are not of power law form), while 
the sum of many flares results in a time averaged spectrum
that can be reasonably well (but not exactly) described
by a power law.

\clearpage

\begin{sidewaystable}
 \caption{\footnotesize Parameters of fits to the cosmic ray proton spectrum.
 Energies are given in TeV, and the normalizations
 $K$ in units (m$^{2}$~s~sr~GeV)$^{-1}$.
 \label{tab:par}}
\begin{center}
\renewcommand{\arraystretch}{1.65}
 
\begin{tabular}{ | c || c | c | c | c | c | c | c || c | }
\hline
~~ & ~~PAMELA~~ & ~~AMS02~~& ~~ATIC ~~ & 
~~CREAM~~ & ~~CALET~~ & ~~DAMPE~~ &
~~NUCLEON ~~ & ~~~All Data~~~ \\ 
\hline 
$\alpha_1$
& $2.850 \pm 0.043$ 
& $2.849^{+0.006}_{-0.005}$
& $2.79^{+0.20}_{-0.04}$ 
& --
& $2.81 \pm 0.01$
& $2.750 \pm 0.005$
& -- 
& $2.80\pm 0.03$ \\ 
$\alpha_2$
& $2.67 \pm 0.06$
& $ 2.716^{+0.037}_{-0.056}$
& $2.62^{+0.02}_{-0.05}$ 
& $2.58 \pm 0.01$
& $2.55 \pm 0.01$
& $2.58 \pm 0.01$
& $2.44^{+0.5}_{-0.3}$
& $2.57^{+0.04}_{-0.06}$ \\ 
$\alpha_3$
& -- & -- & --
& $2.84^{+0.07}_{-0.04}$ & -- & $2.86^{+0.07}_{-0.04}$
& $2.86^{+0.5}_{-0.1}$ 
& $2.87^{+0.15}_{-0.10}$ \\ 
$E_b$
& $0.232^{+0.035}_{-0.030}$
& $0.336^{+0.095}_{-0.052}$
& $0.275^{+0.140}_{-0.180}$
& --
& $0.605^{+0.060}_{-0.050}$
& $0.574^{+0.044}_{-0.037}$
& --
& $0.67^{+0.26}_{-0.15}$ \\ 
$w$
& $0^{+0.25}_{-0}$
& $0.04^{+0.21}_{-0.04}$
& $0.34^{+0.50}_{-0.34}$
& -- & $0.25\pm 0.05$
& $0.35^{+0.14}_{-0.31}$
& -- 
& $0.27\pm 0.19$ \\ 
$E_b^\prime$
& -- & -- & --
& $16^{+5}_{-4}$
& --
& $12.8^{+3.9}_{-2.1}$
& $9.5^{+8.0}_{-2.0}$
& $16^{+13}_{-8}$ \\ 
$w^\prime$
& -- & -- & --
& $0^{+0.5}_{-0}$
& --
& $0.37^{+0.22}_{-0.032}$
& $0.24^{+0.7}_{-0.2}$
& $0.35^{+0.4}_{-0.3}$ \\ 
\; $K_{(0.1\,{\rm TeV})}/10^{-2} \;$
& $4.60 \pm 0.06$
& $4.42 \pm 0.02$
& $4.4^{+0.2}_{-0.4}$
& --
& ~$4.34 \pm 0.01 $~
& ~$5.55 \pm 0.09 $~
& --
& ~$4.40 \pm 0.02 $~ \\ 
\; $K_{(2\,{\rm TeV})}/10^{-5} \;$
& --
& --
& --
& ~$1.22 \pm 0.01 $~
& $0.952^{+0.013}_{-0.015}$
& ~$1.15 \pm 0.02 $~
& ~$1.26 \pm 0.04 $~
& ~$1.02 \pm 0.02 $~ \\ 
\hline
$\chi^2_{\rm min}$
& 1.5
& 3.3
& 1.5
& 1.4
& 17.3
& 6.4
& 3.4 
& 46.5 \\ 
$N_{\rm dof}$
& $16 -5$
& $31 -5$
& $15-5$
& $12-5$
& $23-5$
& $17-8$
& $11-5$
& $124-8$ \\ 
\hline
\end{tabular}
\end{center}
\end{sidewaystable}

\clearpage

\end{document}